\documentclass{osa-article}

\usepackage{diagbox}
\usepackage{xcolor}
\usepackage{soul}

\journal{oe}

\articletype{Research Article}

\begin{document}

\title{Finite-key analysis for twin-field quantum key distribution based on generalized operator dominance condition  }

\author{Rui-Qiang Wang,\authormark{1,2,3} Zhen-Qiang Yin,\authormark{1,2,3,*} , Feng-Yu Lu \authormark{1,2,3},Rong Wang\authormark{1,2,3},Shuang Wang\authormark{1,2,3},Wei Chen\authormark{1,2,3},Wei Huang\authormark{4},Bing-Jie Xu\authormark{4},Guang-Can Guo\authormark{1,2,3}and Zheng-Fu Han \authormark{1,2,3} }

\address{\authormark{1}CAS Key Laboratory of Quantum Information, University of Science and Technology of China, Hefei 230026, P. R. China\\
\authormark{2}Synergetic Innovation Center of Quantum Information $\&$ Quantum Physics, University of Science and Technology of China, Hefei, Anhui 230026, P. R. China\\
\authormark{3}State Key Laboratory of Cryptography, P. O. Box 5159, Beijing 100878, P. R. China \\
\authormark{4} Science and Technology on Communication Security Laboratory,
		Institute of Southwestern Communication, Chengdu, Sichuan 610041, China
}
\email{\authormark{*}yinzq@ustc.edu.cn} 



\begin{abstract}
Quantum key distribution (QKD) can help two distant peers to share secret key bits, whose security is guaranteed by the law of physics. In practice, the secret key rate of a QKD protocol is  always lowered with the increasing of channel distance, which severely limits the applications of QKD. 
	     Recently, twin-field (TF) QKD has been proposed and intensively studied, since it can beat the rate-distance limit and greatly increase the achievable distance of QKD. Remarkalebly, K. Maeda et. al. proposed a simple finite-key analysis for TF-QKD based on operator dominance condition. Although they showed that their method is sufficient to beat the rate-distance limit, their operator dominance condition is not general, i.e. it can be only applied in three decoy states scenarios, which implies that its key rate cannot be increased by introducing more decoy states, and also cannot
	      reach the asymptotic bound even in case of preparing infinite decoy states and optical pulses. Here, to bridge this gap, we propose an improved finite-key analysis of TF-QKD through devising new operator dominance condition. We show that by adding the number of decoy states, the secret key rate can be furtherly improved and approach the asymptotic bound. Our theory can be directly used in TF-QKD experiment to obtain higher secret key rate. Our  results can be directly used in experiments to obtain higher key rates.
\end{abstract}

\section{Introduction}
	Quantum key distribution (QKD) \cite{bennett1984,ekert1991quantum}  provides two
	distant parties (Alice and Bob) a secret string of random bits against any eavesdropper (Eve), who may have unlimited power of computing but is just assumed to obey the law of
	quantum mechanics \cite{scarani2009security,lo2014secure}. During last three decades, QKD
	has been developed rapidly both in theory and experiment.
	In theory, the security of QKD is thoroughly analyzed \cite{scarani2009security}, while a variety of novel protocols, e.g. decoy states \cite{hwang2003quantum,wang2005beating,lo2005decoy} and measurement-device-independent (MDI) protocol \cite{lo2012measurement}, are proposed. 
	In experiment, it is on the way to a wide range of QKD networks \cite{tang2016measurement,wang2014field}, even a satellite-to-ground quantum key distribution has been realized \cite{liao2017satellite}. Among all these above mentioned QKD protocols and experiments, there are some fundamental
	limits \cite{takeoka2014fundamental,pirandola2017fundamental} on the secret key rate versus channel distance. For instance, Pirandola-Laurenza-Ottaviani-Banchi (PLOB) bound $R\leqslant{-log_2(1-\eta)}$ \cite{pirandola2017fundamental} gives the precise limit on the secret key rate $R$ under a given channel transmittance $\eta$ for any repeaterless QKD protocols. 
	
	To surpass the PLOB bound, a possible way is to introduce at least one middle node in the protocol. However, this is not a sufficient condition, i.e. the original MDI-QKD protocol does have a middle node but is still unable to beat the PLOB bound. Indeed, some extensions of MDI-QKD can improve its rate scaling from $\eta$ to $\sqrt{\eta}$ by either using quantum memories \cite{panayi2014memory,abruzzo2014measurement} or quantum non-demolition measurement\cite{azuma2015all}. Albeit these setups can be considered to be the simplest examples of quantum repeaters\cite{sangouard2011quantum,duan2001long} which are the ultimate solution to trust-free long-distance quantum communications \cite{piparo2014long}, quantum memories or quantum non-demolition measurement is quite challenging at present.
	
	Remarkablely, twin-field (TF) QKD protocol, proposed by Lucamarini et al. \cite{ lucamarini2018overcoming}, is capable of overcoming
	PLOB bound without needing quantum memories or quantum non-demolition measurement. TF-QKD, known as a variant of MDI-QKD \cite{lo2012measurement}, uses single-photon click to generate key bit rather than two-photon click in the original MDI-QKD, which is critical for its advantage of beating PLOB bound. Inspired by this dramatic breakthrough, some variants of TF-QKD have  been proposed consequentially \cite{ma2018phase,wang2018twin,curty2019simple,cui2019twin,lin2018simple,yin2019twin}, and some realizations \cite{ minder2019experimental,wang2019beating,liu2019experimental,zhong2019proof, grasselli2019practical} have been reported. 
	
	In Refs.\cite{cui2019twin,curty2019simple,lin2018simple}, authors independently proposed a variant of TF-QKD featuring simpler process and higher key rate, since phase postselection is removed. For simplicity, we call this protocol No-phase-post selection(NPP) TF-QKD in the remainder of the paper. The original papers on NPP-TFQKD \cite{cui2019twin,curty2019simple,lin2018simple} gave security proof based on different methods, but a finite-key analysis was missing. Later, some proofs of NPP-TFQKD on finite-key scenario are proposed\cite{maeda2019repeaterless,lu2019practical}. Remarkably, K.Maeda et. al. proposed a simple finite-key analysis for NPP-TFQKD based on operator dominance condition\cite{maeda2019repeaterless}. Their method is sufficient to beat the rate-distance limit when the amount of pulses in the signal mode sent by Alice and Bob reaches $10^{12}$, which is much smaller than the result obtained in Ref.\cite{lu2019practical}. However, their operator dominance condition is not general which can be only applied in three decoy states scenarios. Hence, one cannot increase its key rate by introducing more decoy states.
	In this work, inspired by the idea of operator inequality, we propose another operator inequality condition which can be applied to any number of decoy states scenarios. This leads to a higher key rate than that of \cite{maeda2019repeaterless}. In section I, we briefly review the flow of NPP-TFQKD and the idea of using operator dominance condition to analyze its security, then propose a new operator inequality. In section II, we present a new operator inequality and a virtual protocol whose security is naturally based on the proposed operator inequality. In section III, we convert the virtual protocol into an actual protocol which is practical in real-life, and a simulation in finite-key case is given then.
	Finally a conclusion is present. 

	\section{ operator dominance condition and virtual protocol}
	  The flow of NPP-TFQKD is sketched  in Fig 1. In order to share security key, Alice and Bob both send optical pulses to Charlie, who controls the untrusted  central  station. Both of Alice and Bob randomly switch among code mode and test mode independently. They use code mode to share keys and test mode to estimate the potential information leakage. 
	
	\begin{figure}[hbt!]
		\centering
		\includegraphics[scale=0.4]{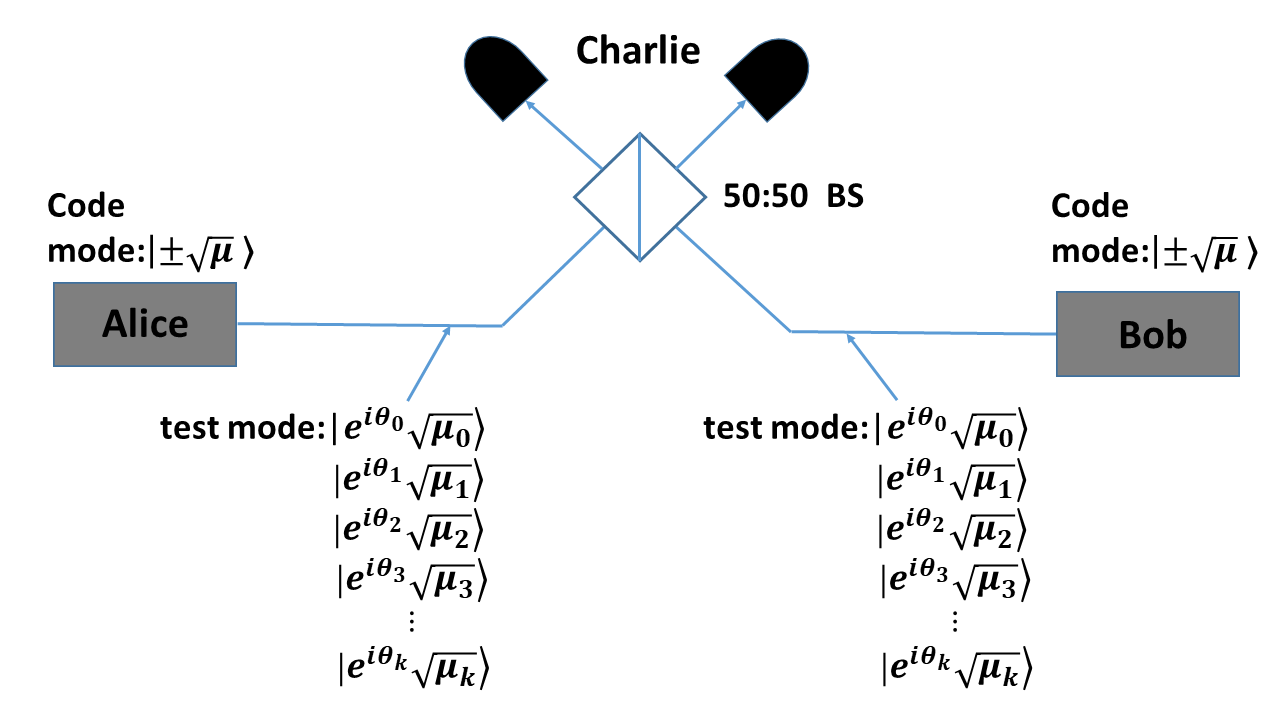}
		\caption{Illustration of NPP-TFQKD protocol, Alice and Bob generate their raw key from the rounds in which they both select the code mode and Charlie declares a successful detection . They encode their key bits in the phase of their coherent states. When the coherent states are in-phase (anti-phase), Charlie's 50:50 beam splitter interference should cause a click in left(right)
		detector. Theses phase-randomized coherent states in the test mode are only used to  monitor the amount of leak.  }
	\end{figure}
	
	In the code mode, Alice and Bob randomly applying 0 or $\pi$ phase shifting to the weak coherent state  $|\sqrt\mu\rangle$. Then they send the pulses to Charlie who measures and announces whether these two quantum states are in-phase or anti-phase when the detection is successful. Bob flips his bit when anti-phase was announced. By this way, they can share random bits. In the test mode, both of the senders randomize the optical phase $\theta$ and switch among several intensities $\{\mu_0,\mu_1,\mu_2,\mu_3,\cdots,\mu_k \}$. They use these phase-randomized coherent states to monitor the amount of information leakage. There are two ways to estimate and generate secret key bits. The first way is to directly calculate Eve's information which is limited by Holevo bound, just like Refs.\cite{yin2019twin,lu2019practical}. The other way is calculating the phase error in an equivalent protocol where  Alice and Bob introduce auxiliary qubits $A$ and $B$, just like Refs \cite{curty2019simple,lin2018simple,maeda2019repeaterless}. It seems that the latter one is better when finite-key effect is considered. Thus, we follow the latter way and introduce the virtual protocol used here.
	
	 Alice and Bob's procedure in each trial of the code mode is equivalently implemented  by preparing  the following joint quantum state
	
	\begin{small}
	 \begin{eqnarray}
	  ( \frac{|0\rangle_A |\sqrt{\mu}\rangle_{C_A}+|1\rangle_A|-\sqrt{\mu}\rangle_{C_A}}{\sqrt{2}} ) 
	  \otimes (\frac{|0\rangle_B|\sqrt{\mu}\rangle_{C_B}+|1\rangle_B|-\sqrt{\mu}\rangle_{C_B}}{\sqrt{2}}), \notag
	 \end{eqnarray}
	\end{small}where $\{|0\rangle ,|1\rangle  \}$ denotes the qubit in $Z$ basis, and $C_A(C_B)$ denotes the optical pulse sent by Alice(Bob). Alice and Bob retain the pairs of $A$ and $B$ in case of Charlie announcing a successful detection. When the number of successful detection is sufficiently large, Alice and Bob measure the qubits $A$ and $B$ in the $Z$ basis to collect sifted key bits. In order to know the information leakage, we have to estimate the phase error rate in Z basis which is equal to the bit error rate in X basis instead of $Z$ basis. This corresponds to the pair in either state $|+\rangle |-\rangle$ or  $|-\rangle |+\rangle$ where $\{|\pm\rangle = (|0\rangle \pm |1\rangle )/\sqrt{2} \}$  denote the qubit in X basis. Hence, the key point is that how we estimate the bit error rate if Alice and Bob virtually measure the retained pairs of $A$ and $B$ with $X$ basis. Supposing that Alice and Bob make the $X$ basis measurement before sending out the optical pulses, we can rewrite the joint quantum state as 
\begin{small}
	\begin{eqnarray}
	(\sqrt{c_+} |+\rangle_A |\sqrt{\mu_{even} } \rangle_{C_A} +\sqrt{c_-}|-\rangle_A |\sqrt{\mu_{odd} }  \rangle_{C_A} )   
	\otimes  
	(\sqrt{c_-}|+\rangle_B|\sqrt{\mu_{odd} }\rangle_{C_B} +\sqrt{c_+} |-\rangle_B|  \sqrt{\mu_{even}}_{C_B}  ) , 
	\end{eqnarray}
\end{small}where $c_+=e^{-\mu}cosh\mu$ and $c_-=e^{-\mu}sinh\mu $. The state $|\sqrt{\mu_{even}}\rangle = (|\sqrt{\mu}\rangle+ |-\sqrt{\mu}\rangle)/2\sqrt{c_-}$ consists of even photon numbers, and the state $|\sqrt{\mu_{odd}}\rangle =(|\sqrt{\mu}\rangle-|-\sqrt{\mu}\rangle)/2\sqrt{c_-} $ consists of odd photon numbers. After tracing out the qubits $A$ and $B$, we can find that 
\begin{small}
	\begin{eqnarray}
	\rho_{C_AC_B}=p_{even}\rho_{even} +p_{odd}\rho_{odd}. 
	\end{eqnarray}
\end{small}Here $ p_{even}=c^2_+ + c^2_-= e^{-2\mu}\cosh2\mu $  and the quantum state $ \rho_{even} $  reads

\begin{small}
	\begin{align}
	p_{even}\rho_{even}=c_+^2|\sqrt{\mu}_{even}\sqrt{\mu}_{even}\rangle\langle    \sqrt{\mu}_{even}\sqrt{\mu}_{even}|     
	+c_-^2|\sqrt{\mu}_{odd}\sqrt{\mu}_{odd}\rangle \langle \sqrt{\mu}_{odd}\sqrt{\mu}_{odd}| ,
	\end{align}
\end{small}where $ p_{odd}=1-p_{even} $ and the quantum state $\rho_{odd}$ reads 
\begin{small}
	\begin{align}
	p_{odd}\rho_{odd} =c_+c_-|\sqrt{\mu}_{even}\sqrt{\mu}_{odd}\rangle\langle \sqrt{\mu}_{even}\sqrt{\mu}_{odd}|    
	+c_-c_+|\sqrt{\mu}_{odd}\sqrt{\mu}_{even}\rangle\langle \sqrt{\mu}_{odd}\sqrt{\mu}_{even}|. \notag 
	\end{align}
\end{small}

Evidently, if Alice and Bob are able to prepare $\rho_{even}$ and $\rho_{odd}$, the security of NPP-TFQKD will be completely equivalent to the original MDI-QKD with single photon source, and then some previous security analyses in finite-key case can be adapted conveniently. However, $\rho_{even}$ and $\rho_{odd}$ are non-classical optical pulses, which are impossible to prepare with off-the-shelf devices. The essential contribution of Ref.\cite{maeda2019repeaterless} is finding an efficient way to approximate $\rho_{even}$ just by using some phase-randomized coherent states. Specifically, they proposed an operator dominance condition which reads

\begin{small}
	\begin{eqnarray}
	p_{0}^2 \tau(0) +p_{1}^2\tau(\mu_1)   
	-\Gamma \tau(\mu_2)  \ge \Lambda \rho_{even}. \notag 
	\end{eqnarray} 
\end{small}
Here, $p^2\tau(\mu) =
p^2\sum_{n,m} \frac{\mu^{n+m} e^{-2\mu}}{n!m!}|n\rangle \langle n| \otimes |m\rangle \langle m|$ corresponds to the joint quantum state in case of Alice and Bob both preparing phase-randomized weak coherent pulses with mean photon-number $\mu$, and the corresponding probability is $p^2$. This operator inequality implies that Alice and Bob's joint phase-randomized weak coherent state can be reinterpreted as a mixture of $\rho_{even}$, weak coherent states with a different intensity, and some "junk" states. Hence, it's possible to bound the yield of $\rho_{even}$ just through preparing phase-randomized weak coherent states with three intensities. However, this inequality is not tight and cannot improved by introducing $4$ or more intensities.

Intuitively, $\rho_{even}$ is just related to the Fock states whose the total photon-number emitted by Alice and Bob is even, thus it is reasonable to devise operator dominance condition with just these even photon-number states. Based on this consideration, we propose another operator dominance condition which reads
\begin{small}
	\begin{eqnarray}
	p_{0}^2   \tau_{\mu_0,even} +p_{1}^2   \tau_{\mu_1,even}   
	-\Gamma \tau_{\mu_2,even} \ge \Lambda \rho_{even},    
	\end{eqnarray}
\end{small}
where the quantum state $\tau_{\mu,even} =  \sum_{k=0}^{\infty} \sum_{j=0}^{2k} \frac{\mu^{2k}e^{-2\mu}}{j!(2k-j)!}   |j\rangle \langle j| \otimes  |2k-j\rangle \langle 2k-j| $ corresponds to Alice and Bob's joint phase-randomized weak coherent state with all odd total photon number states eliminated. The proof of this operator inequality is given in Appendix A. 
     
     To analyze the security of NPP-TFQKD with the proposed operator inequality, we employ the following virtual protocol, whose security can be proved by  Eq (4) easily. \\
     \\
    Step 1: Alice and Bob choose a label from label set
    $ \left\{ "code","0_{even}","0_{odd}","1_{even}","1_{odd}","2_{even}","2_{odd}" \right\}$ with probability $p_c^2, p_0^2  p_{\mu_0,even},  p_0^2 p_{\mu_0,odd},   p_1^2 p_{\mu_1,even},  p_1^2 p_{\mu_1,odd},  p_2^2 p_{\mu_2,even},     p_2^2 p_{\mu_2,odd}$, respectively. According to the label, they perform one of the following procedures.
    
    "code": Alice and Bob generate random key bits $x$ and $y$, send weak coherent states $|(-1)^x \sqrt{\mu}\rangle$ and $|(-1)^y \sqrt{\mu}\rangle$ to Charlie, respectively.
    
    "$0_{even}$": Alice and Bob send a joint quantum state $\tau_{\mu_0,even}$ to Charlie.
    
    "$0_{odd}$": Alice and Bob send a joint quantum state $\tau_{\mu_0,odd}$ to Charlie.
    
    "$1_{even}$": Alice and Bob send a joint quantum state $\tau_{\mu_1,even}$ to Charlie.
    
    "$1_{odd}$": Alice and Bob send a joint quantum state $\tau_{\mu_1,odd}$ to Charlie.
     
    "$2_{even}$": Alice and Bob send a joint quantum state $\tau_{\mu_2,even}$ to Charlie.
     
    "$2_{odd}$": Alice and Bob send a joint quantum state $\tau_{\mu_2,odd}$ to Charlie.
    
  Here, the variable  $p_{\mu,even}$ = $e^{-2\mu}$ $\cosh2\mu $ ($p_{\mu,odd}=e^{-2\mu}$ $\sinh2\mu$ ) denotes that the proportion of $\tau_{\mu,even}$ ($\tau_{\mu,odd}$) in $\tau(\mu)$. \\
    \\
    Step 2: Alice and Bob repeat Step 1 for $N_{tot}$ times.     \\
     \\
    Step 3: Charlie receives the incoming pairs of optical pulses, and announces whether the phase difference was successfully detected for each pair he received. For successful detection, he also announces it was in-phase or anti-phase.  \\
    \\
    Step 4: Let $\gamma_c$ be the number of detected rounds for which both Alice and Bob select label "code". Alice concatenates the random key bits for the $\gamma_c$ rounds to define
    her sifted key. Bob defines his sifted key in the same way except that he flips all the bits for the rounds in which Charlie declared anti-phase detection. Let $\gamma_{0,even}$, $ \gamma_{0,odd}$, $\gamma_{1,even}$, $ \gamma_{1,odd}$,  $\gamma_{2,even}$, $ \gamma_{2,odd}$ be the number of detected rounds for  which both Alice and Bob  send  the  quantum states $\tau_{\mu_0,even},\tau_{\mu_0,odd},\tau_{\mu_1,even},\tau_{\mu_1,odd},\tau_{\mu_2,even},\tau_{\mu_2,odd}$. Let $\gamma_{sum,even}=\gamma_{0,even}+\gamma_{1,even}$. \\
    \\
     Step 5: Alice announces $H_{EC}$ bits of syndrome of a error correction code for her sifted key to perform key reconcilation. Bob reconciles his sifted key accordingly. Alice and Bob verify the correction by comparing $\zeta'$ bits $universal_2$ hashing \cite{carter1979universal}  \\
         \\
    Step 6: They apply the privacy amplification to obtain final keys of length 
    \begin{small}
        \begin{eqnarray}
            G=\gamma_c-(\gamma_c h(f(\gamma_{sum,even},\gamma_{2,even})/\gamma_c )  ) -H_{EC}-\zeta -\zeta',
         \end{eqnarray}
    \end{small}
    where the function $h(x)=-xlog_2x-(1-x)log_2(1-x)$ for $x \le 1/2$ and $h(x)=1$ for x>1/2 and $\zeta$ is related to the security parameter of secret key bits. The function $f(\gamma_{sum,even},\gamma_{2,even})$ is essential for the security, since it gives an upper bound of detection number for Alice and Bob virtually prepare $p_{even}\rho_{even}$ in the $\gamma_c$ sifted key generations rounds. Its definition will be introduced below.

    Define $\gamma_{c,even}$ is the exact detection number for Alice and Bob virtually preparing $p_{even}\rho_{even}$ in the $\gamma_c$ sifted key generations rounds, then $\gamma_{c,even}/\gamma_c$ is just the phase error rate of sifted key bits. We construct  
    a function $f$ subjected to 
          \begin{small}
             \begin{equation}
             Prob\{ \gamma_{c,even} \leq f(\gamma_{sum,even},\gamma_{2,even})  \} \geq 1-\epsilon,
             \end{equation}
          \end{small} which means that $f$ bounds $\gamma_{c,even}$ with a failure probability $\epsilon$. According to Ref.\cite{ scarani2009security}, we will know that this formula implies that the virtual  protocol is $\epsilon_{sec}$-secure where the security parameter $\epsilon_{sec}=\sqrt{2}\sqrt{\epsilon + 2^{-\zeta} } +2^{-\zeta'}$. 
      Now, we start to construct the function $f(\gamma_{sum,even},\gamma_{2,even})$. Since Eq.(4) holds, we can safely suppose that 
      \begin{small}
      	\begin{align}
      	(p_{0}^2+p_{1}^2)\tau_{sum,even} =\Gamma \tau_{\mu_2,even} +\Lambda \rho_{even} +\Delta \rho_{junk} ,
      	\end{align}
      \end{small}
      where 
      \begin{small}
      	\begin{align}
      	(p_0^2+p_1^2) \tau_{sum,even}=p_0^2\tau_{\mu_0,even} +p_1^2\tau_{\mu_1,even}.
      	\end{align}
      \end{small}
      
      We can immediately observe $\gamma_{sum,even}$, as it is the number of detection rounds that Alice and Bob prepare the state $\tau_{sum,even}$, i.e. $\tau_{\mu_0,even}$ or $\tau_{\mu_1,even}$. Besides, since $\tau_{sum,even}$ is a mixture of $\tau_{\mu_2,even}$, $\rho_{even}$ and $\rho_{junk}$, $\gamma_{sum,even}$ is the sum of the numbers of detection rounds for components $\tau_{\mu_2,even}$, $\rho_{even}$, and $\rho_{junk}$, namely $\gamma_{2',even},\gamma_{c',even},\gamma_{junk}$.  Evidently, $\gamma_{2',even}$ is a Bernoulli sampling from a population with $\gamma_{2',even}+\gamma_{2,even}$, since $\tau_{\mu_2,even}$ shares the same density matrix for the rounds that Alice and Bob choose the label "$2_{even}$". Similarly, $\gamma_{c,even}$ is a Bernoulli sampling from a population with $\gamma_{c,even}+ \gamma_{c',even}$. Since we know the value of $\gamma_{sum,even}$ and $\gamma_{2,even}$, by the use of Chernoff bound, we get an lower bound on $\gamma_{2',even}$with a failure probability $\frac{\epsilon}{2}$. Then, the fact that $ \gamma_{sum,even}=\gamma_{2',even}+\gamma_{c',even}+\gamma_{junk}$ leads to an upper bound on $\gamma_{c',even}$. Finally, by using Chernoff bound again, we get an upper bound on $\gamma_{c,even}$ with a failure probability less than $\epsilon$. The upper bound reads

      \begin{small}
      	\begin{eqnarray}
      	&f(\gamma_{sum,even},\gamma_{2,even}) = 
      	\frac{p_0^2p_{even}}{\Lambda}  (   \gamma_{sum,even}     
      	-\frac{\Gamma}{p_2^2p_{\mu_2,even} }\gamma_{2,even}   \\ \notag
      	&+ \nu(\gamma_{sum,even},\gamma_{2,even})\sqrt{ -log(\epsilon/2)  }   )  ,
      	\end{eqnarray}
      \end{small}
      where 
      \begin{small}
      	\begin{eqnarray}
      	& \nu(\gamma_{sum,even},\gamma_{2,even}) \simeq   \frac { \sqrt{2\Gamma(p_2^2p_{\mu_2,even} +\Gamma ) } } {p_2^2 p_{\mu_2,even} }   \sqrt{\gamma_{2,even}}   \\ \notag 
      	&	+ \sqrt{  2(1+  \frac {\Lambda}{p_0^2p_{\mu_2,even} } ) } \sqrt { \gamma_{sum,even}-  \frac{\Gamma}{ p_2^2  p_{\mu_2,even} } \gamma_{2,even}  }.
      	\end{eqnarray}
      \end{small}
      The upper bound $f(\gamma_{sum,even},\gamma_{2,even})$ satisfies  
      \begin{align}
      Prob   \left( \gamma_{c,even} \le f(\gamma_{sum,even},\gamma_{2,even}) \right)   \ge 1-\epsilon, \notag 
      \end{align}
      which implies that the virtual protocol is $\epsilon_{sec}$-secure and $\epsilon_{sec}=\sqrt{2}\sqrt{\epsilon+2^{-\zeta}}+2^{-\zeta'}$.
    
\section{Actual protocol}
 
 We have proved the security of virtual protocol in the last section. However, the above virtual protocol is not practical, since Alice and Bob can never prepare the quantum state $\tau_{\mu,even}$ and $\tau_{\mu,odd}$ in practice. Fortunately, what we care about are the yields of $\tau_{\mu,even}$ and  $\tau_{\mu,odd}$, and we note that the phase randomized coherent state $\tau_\mu$ consists of $\tau_{\mu,even}$ and $\tau_{\mu,odd}$. This implies that one can bound $\gamma_{sum,even}$ and $\gamma_{2,even}$ by the idea of decoy states\cite{hwang2003quantum,wang2005beating,lo2005decoy}, albeit we cannot deterministically prepare $\tau_{\mu,even}$.
Inspired by this consideration, we convert the virtual protocol to an actual protocol below.\\
\\
   Step 1: Alice(Bob) chooses a label from $ \left\{ "code","0","1","2",\cdots,"k" \right\}$ with probabilities $p_c,p_0 ,p_1, p_2,  \cdots, p_k$ respectively. Then, according to the label, Alice(Bob) performs one of the following procedures.
      
      "code": She(He) generates a random key bit $x$($y$) and sends a weak coherent state $|(-1)^x \sqrt{\mu}\rangle$($|(-1)^y \sqrt{\mu}\rangle$) to Charlie.
      
      "0": She(He) sends a phase-randomized weak coherent state with intensity $\mu_0$ to Charlie.
      
      "1": She(He) sends a phase-randomized weak coherent state with intensity $\mu_1$ to Charlie.
      
      "2": She(He) sends a phase-randomized weak coherent state with intensity $\mu_2$ to Charlie.
      \begin{eqnarray}
          \vdots  \notag 
      \end{eqnarray}
      
      "k": She(He) sends a phase-randomized weak coherent state with intensity $\mu_k$ to Charlie.\\
      \\
      Step 2: Alice and Bob repeat Steps 1 for $N_{tot}$ times.     \\
      \\
      Step 3: Charlie receives the incoming pairs of optical pulses, and announces whether the phase difference was successfully detected for each pair he received. For successful detections, he also announces it was in-phase or anti-phase.  \\
      \\
      Step 4: Alice and Bob disclose their label choices. Let $\gamma_c$ be the number of detected rounds for which both Alice and Bob select label "code". Alice concatenates the random key bits for the $\gamma_c$ rounds to define
      her sifted key. Bob defines his sifted key in the same way except that he flips all the bits for the rounds in which Charlie declared anti-phase detections. Let $\gamma_{ij}$ be the number of detected rounds for  which Alice choose the label "i" and Bob choose the label "j".
      \\
      \\
      Step 5: Alice announces $H_{EC}$ bits of syndrome of a error correction code for her sifted key to perform key reconcilation. Bob reconciles his sifted key accordingly. Alice and Bob verify the correction by comparing $\zeta'$ bits $universal_2$ hashing \cite{carter1979universal}  \\
      \\
      Step 6: They apply the privacy amplification to obtain final keys of length 
         \begin{small}
         	\begin{align}
         	G=\gamma_c-(\gamma_c h(f(\overline {\gamma_{sum,even}},\underline{\gamma_{2,even}})/\gamma_c )  ) -H_{EC}-\zeta -\zeta', 
         	\end{align}
         \end{small}
     where $\overline{\gamma_{sum,even}}$ denotes the upper bound on $\gamma_{even}$ and $\underline{\gamma_{2,even}}$ denotes the lower bound on $\gamma_{2,even}$. Evidently,  Eq.(11) is the same as Eq.(5) except that $f( {\gamma_{sum,even}},{\gamma_{2,even}})$ is replaced by $f(\overline {\gamma_{sum,even}},\underline{\gamma_{2,even}})$. Since $f(\overline {\gamma_{sum,even}},\underline{\gamma_{2,even}}) \ge f( {\gamma_{sum,even}},{\gamma_{2,even}})$, the condition 
     \begin{small}
     \begin{align}
     Prob   \left( {\gamma_{c,even} }\le f(\overline {\gamma_{sum,even}},\underline{\gamma_{2,even}}) \right)   \ge 1-\epsilon \notag 
     \end{align}
     \end{small} holds if both $\overline {\gamma_{sum,even}}$ and $\underline{\gamma_{2,even}}$ are correctly estimated. Furtherly defining an extra failure probability of estimation of  $\underline{\gamma_{2,even}} $ and $ \overline{\gamma_{sum,even} }$ as $\varepsilon_{err}$, we conclude that the security parameter of the actual protocol $\epsilon_{sec}=\sqrt{2} \sqrt{\epsilon'+ 2^{-\zeta}} + 2^{-\zeta'}$ and $\epsilon'=\epsilon+\varepsilon_{err} $.
     
     We note that  in some security proofs of QKD protocols, virtual protocol is completely as same as the actual protocol in terms of key bit and Eve’s system. Indeed, we argue that this condition has been met in our proof. Note that in the virtual protocol defined in the main text, to evaluate X-basis error rate, Alice and Bob prepare $\tau_{\mu_0,even},\tau_{\mu_0,odd},\tau_{\mu_1,even},\tau_{\mu_1,odd},\tau_{\mu_2,even},\tau_{\mu_2,odd}$ with probabilities $p_0^2p_{\mu_0,even},p_0^2p_{\mu_0,odd},p_1^2p_{\mu_1,even}, p_1^2p_{\mu_1,odd},p_2^2p_{\mu_2,even},p_2^2p_{\mu_2,even},$respectively. For instance, recall that $p_{\mu_1,even}\tau_{\mu_1,even}+p_{\mu_1,odd}\tau_{\mu_1,odd}=\tau_{\mu_0}$,which means that virtual protocol can be viewed as preparing phase-randomized coherent state $ \tau_{\mu_0}$. As a result we could describe the virtual protocol in an equivalent way, i.e. protocol2, which is Alice and Bob preparing $\tau_{\mu_0},\tau_{\mu_1},\tau_{\mu_2}$with probabilities $p_0^2,p_1^2,p_2^2$ respectively. From the view of Eve, there is no difference between virtual protocol and protocol2. And Alice and Bob’s key bits are also same because the code modes in virtual protocol and protocol2. The only challenge is that Alice and Bob cannot directly observe the clicks of $\tau_{\mu_0,even},\tau_{\mu_0,odd},\tau_{\mu_1,even},\tau_{\mu_1,odd},\tau_{\mu_2,even},\tau_{\mu_2,odd}$ in the protocol2.

     Fortunately, we can resort to decoy states, i.e. introducing$\tau_{\mu_3},\tau_{\mu_4},\cdots$ Note that by far we do not assume that  $p_c^2+p_0^2+p_1^2+p_2^2 =1 $. Thus we could assume Alice and Bob additionally prepare $\tau_{\mu_3},\tau_{\mu_4},\cdots$ with probabilities $p_3^2,p_4^2,\cdots$ in above virtual protocol and protocol2. Now, the protocol2 here is just the actual protocol defined in the main text. Introducing $\tau_{\mu_3},\tau_{\mu_4},\cdots$is obviously useless in the virtual protocol, then we return to the virtual protocol defined in the main text.

    To calculate the final key length, a simple method of computing $\Gamma$ and $\Lambda$ from ($p_{0},p_{1},\mu,\mu_0,\mu_1,\mu_2$) is given in Appendix A. What's more, using linear program, one can get $\overline{\gamma_{sum,even}}$ and $\underline{\gamma_{2,even}}$ with a failure probability no larger than $\varepsilon_{err}$. For simplicity, we just consider how to calculate $\overline{\gamma_{sum,even}}$ and $\underline{\gamma_{2,even}}$ in the case of four test states whose intensities are $\left\{ \mu_0,\mu_1,\mu_2,\mu_3\right\}$.\\

  The method of computing $ \overline{\gamma_{sum,even}}$ and $\underline{\gamma_{2,even}}$ with linear programming is showed below.

 Indeed the variable $\gamma_{sum,even}$ can be written as 
  
  \begin{small}
      \begin{align}
          \gamma_{sum,even}&=\gamma_{0,even}+\gamma_{1,even}=\triangleq\sum_{k=0}^{\infty}(\sum_{j=0}^{2k}N^{\mu_0\mu_0}_{j,2k-j}+N^{\mu_1\mu_1}_{j,2k-j}),\\ \notag 
      \end{align}
  \end{small}

  and the variable $\gamma_{2,even}$ can be written as
  \begin{small}
      \begin{align}
          \gamma_{2,even}\triangleq \sum_{k=0}^{\infty}\sum_{j=0}^{2k}N^{\mu_2\mu_2}_{j,2k-j} 
      \end{align}
  \end{small}
  
where the variable $N^{\mu\mu}_{j,2k-j}$ denote the number of detected events in which the users sent (j,2k-j) photons and both selected intensity $\mu$. For estimating the upper bound of $\gamma_{sum,even}$, we divide this variable into two parts according to the value of k. As for the part where $k\le 2$, which can be denoted as $\sum_{k=0}^{2}\sum_{j=0}^{2k} (N^{\mu_0\mu_0}_{j,2k-j}+N^{\mu_1\mu_1}_{j,2k-j})$ ,its bound can be calculated with the method in Ref.\cite{curty2014finite}  
 Clearly, variables $N^{\mu_a\mu_b}_{j,2k-j}$ for any $\mu_a,\mu_b \in \{\mu_0,\mu_1,\mu_2,\mu_3\}$ provides a random sampling between each other. Besides, these variables must satisfy the constraints $\sum_{k=0}^{k=2}\sum_{j=0}^{2k}N^{\mu_a\mu_b}_{j,2k-j}\leq \gamma_{a,b}$. By these constraints, for the variable  $\sum_{k=0}^{2}\sum_{j=0}^{2k}( N^{\mu_0\mu_0}_{j,2k-j}+N^{\mu_1\mu_1}_{j,2k-j})$,  one can get its upper bound using linear programming listed in the Supplementary Note 2 of Ref.\cite{curty2014finite}. As for the part $k \ge 3$, We use the Eq(34) in Ref.\cite{zhang2017} to get the upper bound of it from the the expected number of transmitted events  $\sum_{k=3}^{\infty}\sum_{j=0}^{2k}N_{tot}(p_0^2 e^{-2\mu_0}\frac{\mu_0^{2k}}{j!(2k-j)!} + p_1^2e^{-2\mu_1}\frac{\mu_1^{2k}}{j!(2k-j)!} )$
 As for the estimation of the lower bound of $\gamma_{2,even}
 $, we also divide it into two parts according to the value of k, for the part where $k \le 2$, we get its lower with the same method as that of $\gamma_{sum,even}$, for the part where $k\ge3$, we set its lower bound as 0. we denote the total failure probability of estimation of $\overline{\gamma_{sum,even}}$ and $\underline{\gamma_{2,even}}$ as $\varepsilon_{err}=2.60e-20$.

Based on the method given above, we simulate the secret key rate $G/N_{tot}$ as a function of  distance $L$ between Alice and 
Bob when the total number of test states is four. The parameters used in simulation are listed below. We set the  intensity $\mu_0$=5e-4 in the test mode and the parameters ($\mu_1,\mu_2,\mu_3,p_c,p_{0},p_{1},p_2,p_3$) are optimized for each distance. The simulate result is listed in table II. Note that we set $\zeta'$=32 which makes the protocol is $\epsilon_{cor} = 2^{-32}$-cor, while setting $\zeta=2^{-69}$, $\epsilon=2^{-69} $ and $\varepsilon_{err}=2.60e-20$ make the protocol is $\epsilon_{sct}=\sqrt{2}\sqrt{(\epsilon+ \varepsilon_{err})+2^{-\zeta} } $-sct. Finally, all these parameters make the protocol to be $\epsilon_{sec}=\epsilon_{cor}+\epsilon_{sct}=4.6084e-10$-sec.  For comparison, we also simulate the secret key rate of Ref.\cite{maeda2019repeaterless} with the same parameters, and present the result in table III. 
\begin{table}[htbp]
	\centering
\begin{tabular}{cccccc}
	\hline 
	  $e_m$ & $p_d$ &  $\xi$(dB/km) & $\eta_d$ &  $f$ & 
	  $\epsilon_{sec}$ \\
     \hline
     0.03 & $10^{-8}$ & 0.2 & 0.3 & 1.1 & 4.6084e-10  \\
     \hline 
    \end{tabular}
    \caption{List of parameters uesd in the numerical simulations. Here, $e_m$ is loss-independent misalignment error. $p_d$ is dark counting probability. $\xi$ is fiber loss. $\eta_d$ denotes detection efficiency. $f$ is error-correction efficiency. $\epsilon_{sec} $ show that actual protocol is $ \epsilon_{sec}$-secure}
  \end{table}

\begin{table}[!htbp]
\centering
 \begin{tabular}{c|c|c|c|c|c|c}
 \hline
 \diagbox{ pulses}{key rate}{distance}& 0 km & 100 km & 200 km&  300 km & 350 km & 400 km \\
 \hline
 1e11& \textcolor[rgb]{1,0,0} {0.0076} &\textcolor[rgb]{1,0,0}{ 4.2085e-4} &4.0147e-6    & 0  &  0& 0\\
 \hline
 1e12& \textcolor[rgb]{1,0,0} {0.0093} & \textcolor[rgb]{1,0,0} {6.464e-4} &   1.9735e-5  & 8.0269e-7& 1.6073e-9 &0 \\
 \hline
 1e13& \textcolor[rgb]{1,0,0} {0.0110} & \textcolor[rgb] {1,0,0} {7.2618e-4}& \textcolor[rgb]{1,0,0} {4.3012e-5}&  \textcolor[rgb]{1,0,0} {5.2706e-6}& \textcolor[rgb]{1,0,0} {8.8916e-7} &\textcolor[rgb]{1,0,0} {8.5783e-8}\\
 \hline
 1e14& \textcolor[rgb]{1,0,0} {0.0161} & \textcolor[rgb]{1,0,0} {8.3757e-4}  & \textcolor[rgb]{1,0,0} {4.8580e-5} &  \textcolor[rgb]{1,0,0} {9.3155e-6} & \textcolor[rgb]{1,0,0} {1.9658e-6}& \textcolor[rgb]{1,0,0} {2.7717e-7}  \\
 \hline
 Inf & \textcolor[rgb]{1,0,0} {0.0505} & \textcolor[rgb]{1,0,0} {0.0024}    & \textcolor[rgb]{1,0,0} {1.9992e-4} & \textcolor[rgb]{1,0,0} {9.0841e-6}   & \textcolor[rgb]{1,0,0} {2.9624e-6} & \textcolor[rgb]{1,0,0} {1.0456e-6} \\
 \hline
 \end{tabular}
 \caption{  The secret key rate (per pulse) computed by our method. The key rates in red  are higher than the corresponding ones in Table III.}
 \end{table}
 
\begin{table}[!htbp]
\centering
 \begin{tabular}{c|c|c|c|c|c|c}
 \hline
 \diagbox{ pulses}{key rate}{distance}& 0 km & 100 km & 200 km&  300 km & 350 km & 400 km \\
 \hline
 1e11& 0.0032141 &  2.0931e-4&  1.2947e-5& 4.8005e-7& 3.255e-8 & 0\\
 \hline
 1e12&  0.003777 & 2.7426e-4& 2.0166e-5& 1.0789e-6& 1.5406e-7& 0 \\
 \hline
 1e13&  0.00417 & 3.236e-4& 2.6343e-5& 1.6959e-6& 3.0752e-7& 1.0724e-8\\
 \hline
 1e14 &0.0044309 & 3.5787e-4& 3.0927e-5& 2.2078e-6& 4.4945e-7& 3.0326e-8  \\
 \hline
 Inf & 0.0063&8.6679e-4 &8.1977e-5& 7.2654e-6  &8.3271e-7  &4.2944e-7 \\
 \hline
 \end{tabular}
 \caption{ The secret key rate (per pulse) computed by the method in 
 Ref.\cite{maeda2019repeaterless}. }
 \end{table}

As shown in table II and III,those key rates in red  in  table II is higher than those in  table III which show that  the secret key rates of our protocol are obviously higher than those of Ref.\cite{maeda2019repeaterless}, if the pulse number $N_{tot}$ is larger than $10^{13}$ or the channel distance is short (typically shorter than $200$km), which corresponds to the cases that the length of sifted key bits is large. The main reason for this is that we need more test states and linear program to estimate more parameters than the the case in Ref.\cite{maeda2019repeaterless}, which leads to our method is more sensitive to statistical fluctuations. 

\section{conclusion}
Inspired by the idea of operator dominance condition, we propose a generalized operator inequality. Unlike the original one which is only applicable in three decoy states case, the proposed method allows that Alice and Bob use any number of decoy states in the test mode to improve the secret key rate. Additionally, since the proposed operator inequality consists of even photon-number states, a more effective approximate of the quantum state $\rho_{even}$ is made. As a result, higher secret key rate in TF-QKD is obtained in both infinite and finite key regions with considerable key length. Our method can be directly adapted  implementations of TF-QKD.
\\

\section*{Acknowledgement}
This work has been supported by the National Key Research And Development Program of China (Grant Nos. 2016YFA0302600), the National Natural Science Foundation of China (Grant Nos. 61822115,
61775207, 61961136004, 61702469, 61771439,  61627820), National Cryptography Development Fund (Grant No.
MMJJ20170120) and Anhui Initiative in Quantum Infor-
mation Technologies.
\section*{Disclosures}
The authors declare no conflicts of interest.

\section*{Appendix}
We construct the operator dominance condition here. Firstly, we give the methods to calculate the value $\Gamma$ and $\Lambda$ from the parameters ($\mu,  \mu_1, \mu_2,p_0,p_{10},p_{11},p_2$). We choose $\Gamma$ and $\Lambda$ subjected to 
\begin{small}
	\begin{eqnarray}
	\frac{\Gamma}{p_{11}^2}=\frac{\mu_1e^{-2\mu_1}} {\mu_2 e^{-2\mu_2} }
	\end{eqnarray}
\end{small}
and 
\begin{small}
	\begin{eqnarray}
	\frac{p_{even} p_{11}^2}{\Lambda} = \frac{ e^{-2\mu} }  { p_{10}^2e^{-2\mu_0} /p_{11}^2  - e^{-2\mu_1}(\mu_1-\mu_2)/\mu_2 }
	+ \frac{e^{-2\mu } }{\mu_1 e^{-2\mu_1}} \sum_{k=1}^{\infty} 
	\frac{ (k+1)\mu^{2k} }{ \mu_1^{2k} -\mu_2^{2k} }
	\end{eqnarray}
	
\end{small}
Additionally, these variables which subjected to 
\begin{small}
	\begin{eqnarray}
	0 <\frac{\mu_1-\mu_2}{\mu_2}< \frac{p_{10}^2e^{-2\mu_0}} { p_{11}^2e^{-2\mu_1} }
	\end{eqnarray}
\end{small}

Next, We will show why  Eq (4) holds. We expand the left hand side of Eq (4) on the Fock basis $ \sum _{k,k'}(q_{k+k'}/k!k^{'}!)|k,k'\rangle \langle k,k'|$

\begin{small}
	\begin{equation}
	q_n =	
	\begin{cases}
	p_{10}^2e^{-2\mu_0}\mu_0^n+p_{11}^2e^{-2\mu_1}\mu_1^n-\Gamma e^{-2\mu_2}\mu_2^n     \quad   & n\ge 2, \quad  n: even \\
	p_{11}^2e^{-2\mu_1}-\Gamma e^{-2\mu_2}+p_{10}^2 e^{-2\mu_0} \quad   & n=0 
	\end{cases}
	\end{equation}
\end{small}

We suppose another variable

\begin{small}   
	\begin{equation}
	q_n' =	
	\begin{cases}
	p_{11}^2e^{-2\mu_1}\mu_1^n-\Gamma e^{-2\mu_2}\mu_2^n     \quad  & n\ge 2, \quad n:even \\
	p_{11}^2e^{-2\mu_1}-\Gamma e^{-2\mu_2} + p_{10}^2 e^{-2\mu_0}  \quad       &  n=0 
	\end{cases}
	\end{equation}
\end{small}

where $q_n =q_n'=0$ when $n= 2k+1$
. For this protocol, we set $\mu_0$ as a constant 
$5e-4$. Hence, we immediately know that $q_n \ge q_n'$ for any $n \ge 0$. Given that Eq (14),we get 

\begin{small}
	\begin{equation}
	q_n'=
	\begin{cases}
	p_{11}^2\mu_1 e^{-2\mu_1}(\mu_1^{n-1}-\mu_2^{n-1}) \ge 0  &n \ge 2, \quad n:even  \\
	p_{10}^2e^{-2\mu_0}-p_{11}^2e^{-2\mu_1}(\mu_1-\mu_2)/\mu_2 >0 & n=0
	\end{cases}
	\end{equation}
\end{small}  

Given  that Eq (17)  ,Eq (15) can be rewritten as
\begin{small}
	\begin{eqnarray}
	\frac{p_{even} }{\Lambda} = \sum_{k=0}^{\infty} \frac{(k+1) \mu^{2k}}{q_{2k}'}
	\end{eqnarray}
\end{small}

Let $\Pi_e= \sum_{k=0}^{\infty} |2k\rangle \langle 2k |$ and $\Pi_o=\sum_{k=0}^{\infty} |2k+1 \rangle \langle 2k+1| $ be projections to the subspace with even and odd photon numbers,respectively. We denote $\Pi_{\alpha \beta}=\Pi_{\alpha} \otimes \Pi_{\beta}( \alpha,\beta =e,o)$. In according to Eq (3), We get 
\begin{small}
	\begin{eqnarray}
	p_{even} \rho_{even} = \Pi_{ee} |\sqrt{\mu},\sqrt{\mu} \rangle \langle \sqrt{\mu},\sqrt{\mu} | \Pi_{ee}    
	+\Pi_{oo} |\sqrt{\mu},\sqrt{\mu} \rangle \langle \sqrt{\mu},\sqrt{\mu} | \Pi_{oo}    
	\end{eqnarray}
\end{small}
Eq(4) is equivalent to the following set of conditions:

\begin{small}
	\begin{equation}
	p_{even}\sum_{k,k':even}\frac {q_{k+k'}}{ k!k^{'}!} |k,k'\rangle \langle k,k' | \ge \Lambda \Pi_{ee} |\sqrt{\mu},\sqrt{\mu} \rangle \langle \sqrt{\mu},\sqrt{\mu} | \Pi_{ee}    
	\end{equation}
\end{small}

\begin{small}
	\begin{equation}
	p_{even}\sum_{k,k':odd}\frac {q_{k+k'}}{ k!k^{'}!} |k,k'\rangle \langle k,k' | \ge \Lambda \Pi_{oo} |\sqrt{\mu},\sqrt{\mu} \rangle \langle \sqrt{\mu},\sqrt{\mu} | \Pi_{oo}    
	\end{equation}
\end{small}
\begin{small}
	\begin{eqnarray}
	q_{k+k'} \ge 0 (k+k':odd). 
	\end{eqnarray}
\end{small}

$q_{k+k'}=0$ when $k+k'$ is  odd leads to Eq (24) is true. Eq (22) is true if 
\begin{small}
	\begin{eqnarray}
	p_{even} \Pi_{ee} \ge \Lambda |\phi_{ee} \rangle \langle \phi_{ee} |   
	\end{eqnarray}
\end{small}
where 
\begin{small}
	\begin{eqnarray}
	|\phi_{ee} \rangle  =\sum_{k,k':even} (\frac{q_{k+k'}}{k!k^{'}!})^{-1/2}|k,k'\rangle \langle k,k' |\sqrt{\mu},\sqrt{\mu} \rangle  
	\end{eqnarray}
\end{small}
It's easy to know that 
\begin{small}
	\begin{eqnarray}
	\langle \phi_{ee} | \phi_{ee} \rangle = \sum_{k,k':even} \frac{e^{-2\mu} u^{  k+k'}}{q_{k+k'} }  <    \frac{p_{even}}{\Lambda}
	\end{eqnarray}
\end{small}
According to Eq (27), we know that Eq (25) is true and so is 
Eq (22). Similarly, for 
\begin{small}
	\begin{eqnarray}
	|\phi_{oo} \rangle = \sum_{k,k':odd} (  \frac{q_{k+k'}}{k!k^{'}!})  |k,k'\rangle \langle k,k' |\sqrt{\mu},\sqrt{\mu} \rangle  
	\end{eqnarray}
\end{small}
we can konw immediately 
\begin{small}
	\begin{eqnarray}
	\langle \phi_{oo} | \phi_{oo} \rangle = e^{-2\mu} \sum_{k=1}^{\infty } \frac{k\mu^{2k}}{q_{2k} } < \frac{p_{even }}{\Lambda }
	\end{eqnarray}
\end{small}

This leads to that Eq (23) is true.
\bibliography{finite_key}
\end{document}